\documentclass[12pt]{article}
\usepackage{amsmath}
\def\pacs{\rightline}

\begin{document}

\begin{center}
\bigskip

{\large \textbf{BELL'S\ INEQUALITIES\ IN 4-DIMENSION\ RIEMAN'S\ SPACE}}

\medskip TIMUR F.~KAMALOV

MEGALLAN ltd., pr. Mira, 180-3,Moscow129366,Russia

Tel./fax 7-095-282-14-44

E-mail: okamalov@chat.ru \& ykamalov@rambler.ru

\end{center}

It is shown that the nature of quantum statistics can study in assumption of
existence of a background of random gravitational fields and waves,
distributed isotropically in the space. This background is capable of
correlating phases of oscillations of identical microobjects. If such a
background of random gravitational fields and waves is considered as hidden
variables. It is shown that the classic physical of Bells observable in the
4-dimensions Rieman's space gives the value matching the experimental data.
The nature of the entanglement states we are study here.

\bigskip

The quantum theory is a statistical one, which at the same time don't study
this statistics nature considering this problem as being beyond its scope.
The quantum theory non-considers the causes of quantum phenomena, it
postulates the classically inexplicable phenomena of quantum microcosm
observed in experiments as its axioms. Such an approach though not
introducing errors does not explain the experimentally observable phenomena
leaving these incomprehensible from the classical viewpoint and giving rise
to all sorts of paradoxes. On all questions WHY about quantum axiomatic this
theory answer THIS\ IS\ AXIOM. All this mean, that in Classical Physics Wiew
this theory absent axiomatic and can named as Method\ of Indirect
Computations.

While the years have passed from the moment of creation of the quantum
theory, the discussion began in times of its creation has never ceased
throughout this period whether former is complete and whether the hidden
parameters do exist in it. Quite famous is Feynmanns phrase that ''nobody
understands the quantum mechanics, and those who pretend they do - lie''. It
is true that there is no classic causality and classic logic in quantum
mechanics, this lack being compensated by whatever is denoted by the terms
''quantum causality'' and ''quantum logic''. But where does the distinction
of classic from quantum lie? Are classic causality and classic logic absent
in the theory only or in nature as well? Is the reason that an advocate of
the classic logic could be easily blamed lacking abstract thought? Lacking
of classic causality and classic logic in the theory does not imply their
absence in the nature.

Here we will discuss a physical model with random, isotropically and
statistical gravity background, illustrating the hidden variables.

Now, let us try to single out the basic classically incomprehensible
concepts of the quantum theory. First, it is the wave/particle dualism.
Taking into consideration the above mentioned, a particle could acquire wave
properties provided its being influenced by the wave background. Second, it
is Heisenberg's principle of uncertainty. Provided superposition of
non-removable background onto a measurement, it would be impossible to
measure the values precisely. Third, it is the energy balance in an atom.
From the classical physics viewpoint, an electron moving in the electric
field of the nucleus should emit electromagnetic radiation. Can we assume
the background of the whole spectrum of frequencies to impart energy to the
electron, the latter re-emitting it, then the energy balance in the atom
could be offset?

We can to complete the quantum theory with hidden variables without altering
the mathematical apparatus of the quantum mechanics. Are we resulting a
comprehensible theory?

Here we consider Bell's Inequalities in 4-dimensions Rieman's space. The
issue of the necessity to complete the quantum theory has been first
considered in the study by A. Einstein, B. Podolsky, N. Rosen (hereinafter,
EPR) [1]. Let us consider the EPR effect. Two particles, $P$ and $Q$, at the
initial moment interact and then scatter in opposite directions. Let the
first of them be described by the wavefunction $\psi _{P}$, the other by $%
\psi _{Q}$. The system of the two particles $P$ and $Q$ is described by the
wavefunction $\psi _{PQ}$. With this, $\psi _{P}\neq \psi _{PQ}$, $\psi
_{Q}\neq \psi _{PQ}$, $\psi _{PQ}\neq \psi _{P}\psi _{Q}$. Or, $P_{PQ}\neq
P_{P}P_{Q}$. For independent events $P_{P}$ and $P_{Q}$, according to
probability theory, $P_{PQ}=P_{P}P_{Q}$. Where could the dependence of the
object $P$ on the object $Q$ and vice versa originate from, these objects $P$
and $Q$ being considered as distant and non--interacting? The authors EPR
came to the conclusion on incompleteness of the quantum--mechanical
description. To solve this contradiction, an idea has been put forward in
[1] on existence of hidden variables that would make it possible to
interpret consistently the results of the experiments without altering the
mathematical apparatus of quantum mechanics.

Later, it has been proved by von Neumann [2] that quantum mechanics
axiomatic does not allow introduction of hidden variables. It is, however,
important that the argument presented in [2] would not hold valid in certain
cases, e. e.g., for pairwise observable micro objects (for Hilbert space
with pairwise commutable operators) [3]. In 1964, J. S. Bell [4] has
formulated the experimental criterion enabling to decide, within the
framework of the problem statement [1], on the existence of hidden
variables. The essence of the experiment proposed by Bell is as follows.

Let us consider the experimental scheme of EPR. Let there be two photons
that can have orthogonal polarizations $A$ and $B$ or $A^{^{\prime }}$ and $%
B^{^{\prime }}$, respectively. Let us denote the probability of observation
of the pair of photons with polarizations $A$ and $B$ as $\psi _{AB}^{2}$.
Bell has introduced the quantity$\left| \left\langle S\right\rangle \right| =%
\frac{1}{2}\left| \psi _{AB}^{2}+\psi _{A^{^{\prime }}B}^{2}+\psi
_{AB^{^{\prime }}}^{2}-\psi _{A^{^{\prime }}B^{^{\prime }}}^{2}\right| $,
called the Bell's observable; it has been shown that if the local hidden
variables do exist, then $\left| \left\langle S\right\rangle \right| \leq 1$%
. The possibility of experimental verification of actual existence of local
hidden variables has been demonstrated in [4]. The above inequality are
Bell's inequalities. The series of experiments has shown that there does not
exist any experimental evidence of the idea of existence of local hidden
variables as yet, and the existing theories comprising hidden variables are
indistinguishable experimentally. In quantum theories with hidden variables
the wavefunction $\psi =\psi \left( \lambda _{i}\right) $ being the function
of hidden variables $\lambda _{i}$.

Let us consider a physical model of 4-dimension Reiman's space with gravity
background (i. e. the background of gravity fields and waves).We can
illustrate hidden variables as gravity background [5-10].

This is only one of possible versions. We could consider as hidden
variables, for example, the electromagnetic background. We shall not discuss
here the reasons for this version being unfounded, and we shall not consider
it in the present study.

So, let us illustrate the gravitational background as hidden variables. The
gravitational background could be considered negligible and not affecting
the behavior of quantum microobjects. Let us verify whether this is correct.
The quantitative assessments of the gravitational background influence onto
quantum microobjects behavior have not been performed due to the former
having never been examined. The quantum effects are small as well, but their
quantitative limits are known and are determined by the Heisenberg
inequalities. Let us demonstrate the gravitational background being random
and isotropic to affect the phases of microobjects separated in the space
and not interacting. Then we shall be capable of calculating the correlation
factor for these microobjects, hence, the Bell's observable $S$. Having
determined the upper limit for $S$, we shall get the refined Bell's
inequalities taking into consideration the influence of the gravitational
background. Comparing these with the experimental data for the Bell's
observable, we could verify correctness of our approach.

By now, hundreds of experimental studies have been performed on measurement
of the Bell's observable. It can be positively stated that the experimental
value of the Bell's observable has been determined to comply with the
expression $\left| \left\langle S\right\rangle \right| \leq \sqrt{2}$.

Relative oscillations $\ell ^{i}$, $i=0,1,2,3$ of two particles in gravity
fields are described by deviation equations

\begin{center}
\bigskip $\frac{D^{2}}{D\tau ^{2}}\ell ^{i}=R_{kmn}^{i}\ell ^{m}\frac{dx^{k}%
}{d\tau }\frac{dx^{n}}{d\tau }$,
\end{center}

where $R_{kmn}^{i}$ - is the Rieman's tensor.

In this particular case the deviation equations are converted into
oscillation equations for two particles: 
\[
\stackrel{..}{\ell }^{1}+c^{2}R_{010}^{1}\ell ^{1}=0,\quad \omega =c\sqrt{%
R_{010}^{1}}. 
\]

It should be noted that relative oscillations of micro objects $P$ and $Q$
do not depend on the masses of these, but rather on Riemann tensor of the
gravity field. This is important, as in the microcosm we deal with small
masses. Taking into consideration the gravity background, the micro objects $%
A$ and $B$ shall be correlated. It is essential that in compliance with the
gravity theory the deviation equation does only make sense for two objects,
and it is senseless to consider a single object. Therefore, the gravity
background complements the quantum--mechanical description and illustrating
the hidden variables. On the other hand, the von Neumann theorem on
impossibility of hidden variables introduction into quantum mechanics is not
applicable for pairwise commuting quantities (Gudder's theorem [3]).
Introduction of hidden variables in the space with pairwise commuting
operators is appropriate.

The solution of the above equation has the form $\ell ^{1}=\ell _{0}\exp
(k_{a}x^{a}+i\omega t)$, $a=1,2,3$. Here we assume the gravity background to
have a random nature and should be described, similarly to
quantum--mechanical quantities, with probabilistic observations.

Now we can study the nature of the entanglement states. In our assumptions,
there is Background Gravity, which can influence to behavior of the
particles in microcosm. This influence is small one, but can correlated its
the phases, as follow from General Relativity. Each gravity field or wave
with the index $n$ and Riemann tensor $R(n)$ and random phase $\Phi
(n)=\omega (n)t=c\sqrt{R_{010}^{1}(n)}t$, should be matched by a quantity $%
\ell ^{i}(n)$. Therefore, taking into account the gravity background, i. e.
the background of gravity fields and waves, the particles take on properties
described by $\ell ^{i}(n)$.

We will only consider in the present study the gravity fields and waves
which are so small that alter the variables of micro objects $\Delta x$ and $%
\Delta p$ beyond the Heisenberg inequality $\Delta x\Delta p\geq h$. Strong
fields are adequately enough described by the classical gravity theory, so
we do not consider these in the present study. Let us emphasize that the
assumption on existence of such a negligibly small background is quite
natural. With this, we assume the gravity background to be isotropically
distributed over the space.

Regarding the quantum microobjects in the curved space, we must take into
account the scalar product of the two 4-dimensional vectors $A^{\mu }$ and $%
B^{\nu }$ being $g_{\mu \nu }A^{\mu }B^{\nu }$, where for weak gravitational
fields it is possible to employ the value $h_{\mu \nu }$, which is the
solution of Einstein equations for the case of weak gravitational field in
harmonic coordinates and having the form: 
\[
\begin{array}{c}
h_{\mu \nu }=e_{\mu \nu }\exp (ik_{\gamma }x^{\gamma })+\qquad \\[2mm] 
\qquad +e^{\ast }\exp (-ik_{\gamma }x^{\gamma }),
\end{array}
\]
\[
\ g_{\mu \nu }=\eta _{\mu \nu }+h_{\mu \nu }, 
\]

where the value $h_{\mu \nu }<<1$ - is the metrics disturbance, $\eta _{\mu
\nu }$ - metrics Minkovsky's and $e_{\mu \nu }$- the polarization.
Therefore, we shall consider the hidden variables $h_{\mu \nu }$ being the
disturbances of the metrics as distributed in the space with the yet unknown
distribution function $\rho =\rho (h_{\mu \nu })$. Hereinafter, the indices $%
\mu ,\nu ,\gamma $ possess values 0,1, 2, 3.

Then the coefficient of correlation $M$ of projections of hidden variables
unit vector$\ \lambda ^{i}$ onto directions $A^{k}$ and $B^{n}$ specified by
the polarizer in 4-dimension Rieman's space is

\begin{eqnarray*}
\left| M\right|  &=&\left| \left\langle AB\right\rangle \right| =\left|
\langle \lambda ^{i}A^{k}g_{ik}\lambda ^{m}B^{n}g_{mn}\rangle \right| = \\
&=&\left| k\int_{0}^{2\pi }\cos \alpha \cos \left( \alpha +\theta \right)
d\alpha \right| =\left| \cos \theta \right| \text{,} \\
k &=&\frac{1}{\pi },\ \text{becouse }\left| M\right| \leq 1,\text{ } \\
\text{ cos}\alpha  &=&\frac{g_{ik}\lambda ^{i}A^{k}}{\sqrt{\lambda
^{l}\lambda _{l}}\sqrt{A^{l}A_{l}}}
\end{eqnarray*}

here $i,k,m,n$ possess 0,1,2,3; $\theta $ is angle between polarizations; $%
\alpha $ is angle between hidden variables unit vector $\lambda ^{i}$ and $%
A^{k}$.

Then, the Bell's observable $S$ in 4-dimensions space for $\theta =\frac{\pi 
}{4}$, we obtain the maximum value of

$\langle S\rangle =\frac{1}{2}[\langle AB\rangle +\left\langle A^{^{\prime
}}B\right\rangle +\left\langle AB^{^{\prime }}\right\rangle -\left\langle
A^{^{\prime }}B^{^{\prime }}\right\rangle ]=$

$=\frac{1}{2}[\cos (-\frac{\pi }{4})+\cos (\frac{\pi }{4})+\cos (\frac{\pi }{%
4})-\cos (\frac{3\pi }{4})]=\sqrt{2}$,

which agrees fairly with the experimental data. The Bell-type inequality in
4-dimensions space shall take on the form ${\left| \left\langle
S\right\rangle \right| \leq \sqrt{2}}$.

Therefore, we have shown that the classical physics with the gravitational
background gives the value of the Bell's observable matching both the
experimental data and the quantum mechanical value of the Bell's observable.
To sum it up, the description of microobjects by the classical physics
accounting for the effects brought about by the gravitational background is
equivalent to the quantum mechanical descriptions, both agreeing with the
experimental data. From the experiment viewpoint, both of these descriptions
are equivalent; however, employing the quantum mechanical descriptions
demands using the quantum mechanical axioms.
Here we show, that the nature
of the entanglement states can be illustrated by Gravity Background, i.e.
the random and isotropicaly distributing in the space background of gravity
fields and waves. In addisions we was calculate the coefficient of
correlation entanglement states.

\end{document}